%% file: 00main.tex
\documentclass[nonacm, acmlarge, screen]{acmart}
\AtBeginDocument{%
  \providecommand\BibTeX{{%
    \normalfont B\kern-0.5em{\scshape i\kern-0.25em b}\kern-0.8em\TeX}}}
\newcommand{\inlinequote}[2]{
    \emph{``#1''} (\textit{I#2})
}

\newcommand\citetodo[1]{\textcolor{olive}{[CITE]}}

\usepackage[utf8]{inputenc}
\usepackage{enumitem}
\usepackage{csquotes}

\usepackage[para]{footmisc}

\copyrightyear{2024}
\acmYear{2024}
\setcopyright{rightsretained}
\acmConference[CHI '24 Workshop]{CHI '24 Workshop}{May 12, 2024}{Honolulu, Hawai`i, USA}
\acmDOI{}
\acmISBN{}

\begin{document}
\fancyhead[RO]{\fontfamily{LinuxBiolinumT-TLF}\fontsize{8}{10}\selectfont CHI '24 Workshop, May 12, 2024, Honolulu, Hawai`i, USA}

\fancyhead[LE]{\fontfamily{LinuxBiolinumT-TLF}\fontsize{8}{10}\selectfont LLMs as Research Tools Workshop at CHI conference}

\title{Student Reflections on Self-Initiated GenAI Use in HCI Education}

\author{Hauke Sandhaus}
\email{hgs52@cornell.edu}
\orcid{0000-0002-4169-0197}
\affiliation{%
    \institution{Cornell University, Cornell Tech}
    \streetaddress{2 West Loop Rd}
    \city{New York}
    \state{New York}
    \country{USA}
    \postcode{10044}
}

\author{Maria Teresa Parreira}
\orcid{}
\email{mb2554@cornell.edu}
\affiliation{%
    \institution{Cornell University, Cornell Tech}
    \streetaddress{2 West Loop Rd}
    \city{New York}
    \state{New York}
    \country{USA}
    \postcode{10044}
}

\author{Wendy Ju}
\orcid{0000-0002-3119-611X}
\email{wendyju@cornell.edu}
\affiliation{%
  \institution{Cornell Tech}
  \streetaddress{2 West Loop Rd}
  \city{New York}
  \state{New York}
  \country{USA}
  \postcode{10044}
}

\renewcommand{\shortauthors}{Sandhaus, Parreira, and Ju}

\begin{abstract}
This study explores students' self-initiated use of  Generative Artificial Intelligence (GenAI) tools in an interactive systems design class. Through 12 group interviews, students revealed the dual nature of GenAI in (1) stimulating creativity and (2) speeding up design iterations, alongside concerns over its potential to cause shallow learning and reliance. GenAI's benefits were pronounced in the execution phase of design, aiding rapid prototyping and ideation, while its use in initial insight generation posed risks to depth and reflective practice. This reflection highlights the complex role of GenAI in Human-Computer Interaction education, emphasizing the need for balanced integration to leverage its advantages without compromising fundamental learning outcomes.
\end{abstract}


\received{9 March 2024}
\received[revised]{1 May 2024}

\maketitle

\input{01intro}

\input{02related_work}

\input{03method}
\input{04results}

\input{05conclusion}

\begin{acks}
We thank all students for sharing their experience using GenAI in the fall class of interactive device design 2023 at Cornell Tech. 
\end{acks}

\bibliographystyle{ACM-Reference-Format}

\bibliography{GenAI-prototyping-class}

\end{document}

%% file: 01intro.tex
\section{Introduction}

Following a Fall 2023 graduate-level university course on interactive device design, we interviewed student groups in the course about their use of GenAI tools for coursework. Amidst the advent of advanced GenAI tools like GPT-4 and DALL-E, the students were told that they were permitted but not encouraged to use these novel tools in their coursework, provided that they honestly document their usage of these tools as they would document the use of instructions from Youtube or software libraries on GitHub. This class consisted of 6 device design labs and 1 final project\footnote{\hyperlink{https://github.com/FAR-Lab/Interactive-Lab-Hub/tree/Fall2023}{https://github.com/FAR-Lab/Interactive-Lab-Hub}}; students spent their time almost solely on project work by defining and executing design ideas for interactive devices. Post-facto interviews of students' self-initiated use of GenAI tools can help us understand how such tools might be used for good or ill in the context of HCI education. 

In this workshop paper, we present our preliminary findings from our interviews. We found that students engaged with GenAI throughout the entire design process. Students reported it to be most useful in the execution phase of the design process; it helped students by enabling faster iterations and diversifying the perspectives the students considered. 
In contrast, GenAI was less fruitful in the definition stage of the design process. 
Students reflected that GenAI did not help them think critically about their approach and choices, such as user interview script questions; this suggests a risk of diminished learning depth and reduced engagement in reflective thinking. The students' reflections on these dynamics offer insights into how GenAI can both empower and handicap the pedagogical goals of HCI education.

%% file: 02related_work.tex
\subsection{Related Work}



GenAI's capabilities, including producing creative content such as images, texts, and code, present both opportunities and challenges in HCI. Recent studies and workshops highlight the potential for generative AI to facilitate creative human-technology collaboration~\cite{muller_genaichi_2023, shin_integrating_2023}, improve interaction design~\cite{shi_hci-centric_2024}, and support qualitative analysis~\cite{xiao_supporting_2023}. 

The integration of AI tools in educational settings has been explored in terms of both benefits and challenges. Research indicates that technology generally can enhance learning experiences~\cite{saye_technology_1997}; literature on the impact of newer technology, specifically GenAI's impact on programming, starts to get studied~\cite{zheng_chatgpt_2023, amoozadeh_trust_2024}. 

The impact of AI on software development is multifaceted, affecting coding practices, collaboration, and the role of data scientists, showing promise in accelerated programming~\cite{wang_human-ai_2019, schmidt_speeding_2023}.

The broader societal implications of AI, particularly in terms of ethics, bias, and employment, are difficult to anticipate. Design futuring activities~\cite{huang_future_2023, blythe_artificial_2023}, but little empirical work~\cite{baldassarre_social_2023} project AI's broader impacts, and reflect on the need for responsible AI integration, addressing biases, and considering the socio-economic effects of AI adoption.

Due to the novelty of the technology and its rapid progress, many aspects of it are severely understudied by researchers, including those in HCI~\cite{antona_special_2023}. HCI is a field equipped to react to fast-paced technology changes, yet AI itself is difficult to design for with a user-centered design process\cite{yang_re-examining_2020}.

%% file: 03method.tex
\section{Method}
To learn about students' use of GenAI in HCI education, we interviewed 17 students in 12 interviews during Spring 2024. The study recruitment occurred after the students had already received their final grades in the course. 
Two researchers who had been teaching assistants in the course conducted interviews. The semi-structured interviews questioned what students used GenAI for (\autoref{ssec:what}), how students used GenAI~(\autoref{ssec:how}), and finally, had students reflect on GenAI's implications on user-centered design, prototyping and education~(\autoref{sec:why}). 
The students were assured that their interview data would be anonymized in publications and presentations. 
This analysis presents preliminary insights directly from the student's reflections on their GenAI use. 

This class followed a practice-oriented lab structure, where groups of students iterated through the whole process of designing functional prototypes in two-week sprints. At the time, the university had not established any policies regarding the use of GenAI in academic settings. As these technologies were only becoming widely available, there was minimal guidance on integrating them into classes and little guidance on using them within the context of user-centered design. Consequently, all student engagements with this evolving technology were exploratory in nature. 

%% file: 04results.tex
\section{Results}
\subsection{How did students use GenAI?}\label{ssec:how}

Students began with various assumptions about what GenAI \textit{would be} good at. 
While many teams were enthusiastic about the potential of GenAI, one group found it to be generally unhelpful early on and discontinued its use for interactive device design. Some teams shifted to using GenAI only for isolated use cases where they felt it excelled, while a few teams integrated GenAI into the core of their project work. They defaulted to using it when beginning to conceptualize, when they encountered obstacles, and as they reflected on their work. These teams embraced GenAI as an additional `collaborator,` even anthropomorphizing the AI and fully integrating it into the team;
\inlinequote{We would kind of just say - hey let's ask Chatty.}{1}, as well as splitting a hypothetical project price equally with it \inlinequote{Yeah, it would be equal. If we're 5 members and one of them is GPT, it probably get a \$200 share.}{3}.

\subsection{What Are HCI Students Using GenAI for?}\label{ssec:what}
\begin{figure}
    \centering
    \includegraphics[width=0.8\linewidth]{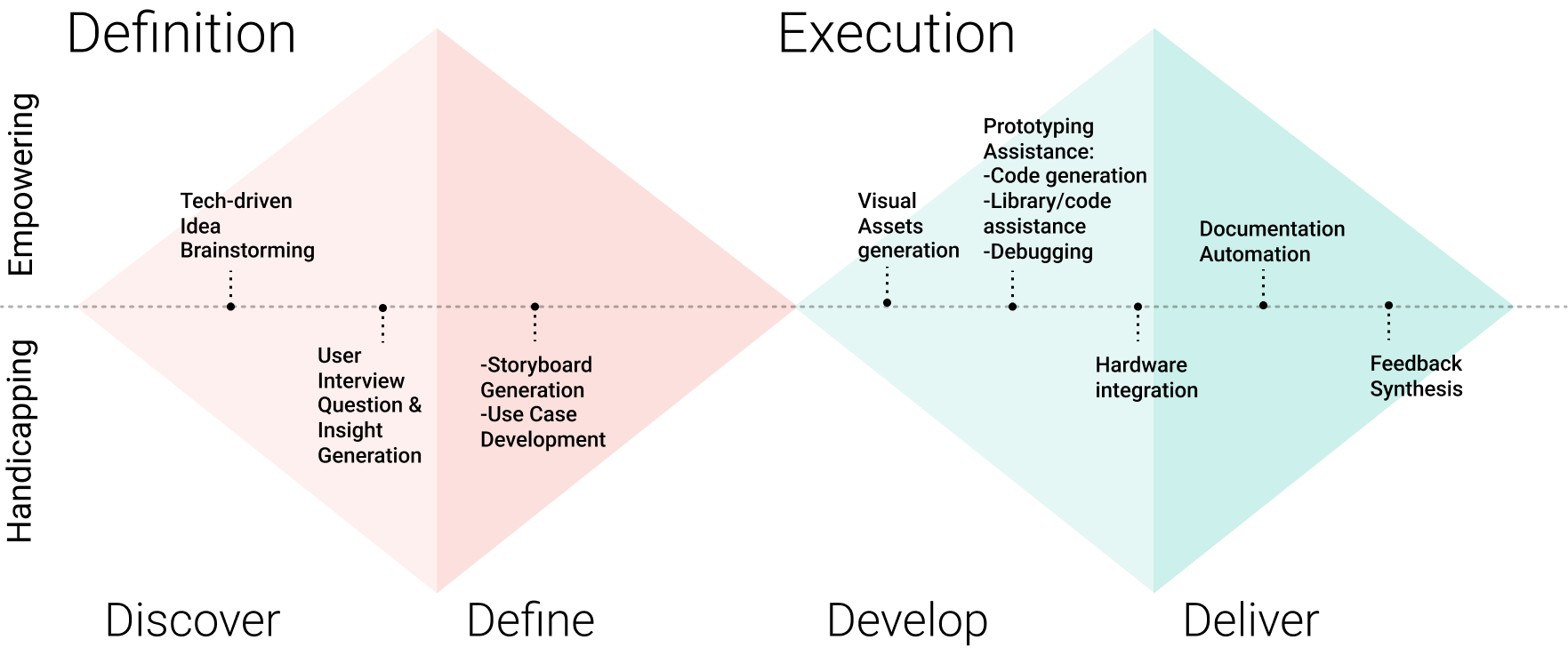}
    \caption{Student's use of generative artificial intelligence mapped to the Double Diamond.}
    \label{fig:double-diamond}
    \Description{This diagram maps the student's use of generative artificial intelligence across the Double Diamond design process model, which consists of four phases: Discover, Define, Develop, and Deliver. The model is visualized with two overlapping diamonds, each divided into two parts. The left diamond, shaded in red and labeled 'Definition,' includes phases Discover and Define, focusing on empowering tasks like tech-driven idea brainstorming, user interview questions, and insight generation. The right diamond, shaded in green and labeled 'Execution,' covers the Develop and Deliver phases, detailing execution tasks like prototyping assistance, visual assets generation, documentation automation, and hardware integration. Dotted lines across the diamonds suggest the iterative nature of the process.}
    \vspace{-0.4cm} 
\end{figure}

Students reported experimenting with various GenAI tools for text generation, such as Grammarly AI\footnote{\href{https://www.grammarly.com/ai}{grammarly.com/ai}}, Google Bard\footnote{\href{https://bard.google.com/}{bard.google.com}}, and Bing Chat\footnote{\href{https://chat.bing.com/}{chat.bing.com}}, as well as for image generation, including Midjourney\footnote{\href{https://www.midjourney.com/}{midjourney.com}}, Google Labs AutoDraw\footnote{\href{https://www.autodraw.com/}{autodraw.com}}, and DALL-E\footnote{\href{https://openai.com/dall-e-2}{openai.com/dall-e-2}}. However, ChatGPT versions 3.5 and 4.0\footnote{\mbox{\href{https://openai.com/research/gpt-4}{openai.com/research/gpt-4}}} were by far the most frequently used tools, with few competitors.
Based on our group interviews, we identified several applications of GenAI throughout the Double Diamond design process\footnote{\href{https://www.designcouncil.org.uk/double-diamond/}{designcouncil.org.uk/double-diamond}} (see \autoref{fig:double-diamond}).

    \begin{itemize}[leftmargin=*,itemsep=0pt,parsep=1pt, label={$\diamond$}]   
    \item \textbf{Discover} 
    -- \textit{Tech-Driven Idea Brainstorming:} GenAI was leveraged for brainstorming, generating ideas for technology applications, and exploring new concepts based on current tech trends.
    
    -- \textit{User Interview Question Generation and Insight Generation:} Students utilized GenAI to come up with questions for user interviews and to summarize and generate insights from user research.
    
    \item \textbf{Define}
    --\textit{Storyboard Generation:} Students used GenAI to create storyboards, visualizing user scenarios, journeys, or interactions, aiding in the definition and planning of the UX/UI design process.
    
    --\textit{Use Case Development:} GenAI assisted in developing user personas and synthesizing user characteristics, needs, and goals.
    \item \textbf{Develop}
    --\textit{Visual Asset Creation:} Tools like DALL-E were utilized for creating visual assets or mockups, supporting the design phase by generating images or icons.
    
    --\textit{Prototyping Assistance:} Tools like ChatGPT were used for coding prototypes, from generating initial code snippets to providing debugging solutions and refining prototypes.
   
    --\textit{Hardware Integration:} Students used GenAI to solve hardware problems, including sensor issues, cabling, advice on hardware libraries, and 3D printing.
    \item \textbf{Deliver}
    --\textit{Documentation Automation:} GenAI was used for automating the creation of project documentation by cleaning up text, expanding notes, and adding code comments.
    
    --\textit{Feedback Synthesis:} GenAI was applied to synthesize and analyze user feedback or test results. Students also used it to reflect on their own work process, as well as other students' work.
    \end{itemize}

This overview, derived from the interviews and mapped to the Double Diamond Design Process (see \autoref{fig:double-diamond}), details the specific ways in which students applied GenAI throughout the user research and design process. It highlights how GenAI facilitated creativity and ideation in the Discover phase and supported documentation and feedback analysis in the Deliver phase.

\subsection{Challenges and Benefits Through the Students' Lens}\label{sec:why}

While students were positive about the technology in many regards, they also appeared wary of GenAI. There was no interview in which students did not mention the risks of the technology, particularly in relation to education and deepness of thought. In the following, we report students' own reflections on the risks and opportunities of the technology. 

\subsubsection{Empowering Aspects of GenAI Use}
GenAI significantly enhances the HCI design education experience in several areas:

\vspace{-0.2cm} \paragraph{Tech-driven Idea Brainstorming} While many students questioned the novelty of GenAI's ideas, it can act as a starting point for creativity, actively encouraging students to raise the bar of what is a \textit{new} idea.  \inlinequote{GPT is basically serving as a baseline on how I would judge others ideas. ... You need to beat the AI.}{6}

\vspace{-0.2cm} \paragraph{Visual Assets Generation} While the visual quality and degree of control of GenAI images are evolving, students found that for creating assets and illustrating unique prototyping and design needs, GenAI can accelerate the iteration process. 

\vspace{-0.2cm} \paragraph{Prototyping Assistance} 
Students reported to be \emph{empowered} by using GenAI for prototyping.
--\textit{Code Generation:} Speeding up the development process, students highlighted how GenAI \inlinequote{allows you to be faster with the prototyping and the execution, and then you have more time to think about
what you want to do.}{5}, allowing more critical thinking and refinement.
--\textit{Library/Code Assistance and Debugging Assistance:} By suggesting relevant libraries, coding practices, and offering debugging assistance, GenAI reduces troubleshooting time as described in many interviews, \inlinequote{I would just ask the questions ... rather than start to google and then try and find stack overflow or community answer questions.}{8}
\vspace{-0.2cm} \paragraph{Documentation Automation:} In the deliver phase, GenAI helped to expand notes into understandable text, assist non-native speakers with translations and write code comments. 

\subsubsection{Handicapping Aspects of GenAI Use}
Conversely, certain applications of GenAI may hinder the depth and authenticity of the HCI design process:

\vspace{-0.2cm} \paragraph{User Interview Question Generation and Insight Generation:} While the interactive device design class did not have an interview focus, students reflected on other classes where they used GenAI for that part. Generated questions might result in less tailored questions, missing deeper insights and the analysis risks oversimplification, leading to superficial understanding. \inlinequote{Now you're learning less in HCI class. ... To code up an interview script, if you don't have ChatGPT you need to read over the script.  ... Everyone in the class, they just dump the script into ChatGPT and try to generate a result.}{5}


\vspace{-0.2cm} \paragraph{Storyboard Generation} At the onset of the class, students experimented with employing GenAI for storyboard ideation. However, they found that the storyboards generated by GenAI were unhelpful, prompting a return to traditional sketching with pen and paper. The use of GenAI resulted in time being devoted to attempting to generate desired outcomes rather than facilitating quick thinking and iteration. It also led to the production of generic storyboards that lacked depth and a personal touch. \inlinequote{Like, physically as I'm drawing, I'm imagining,... Like, why am I placing it here? Why am I drawing it here? Why am I drawing it in this shape?}{9}


\vspace{-0.2cm} \paragraph{Hardware integration} 
Students reported that GenAI was not yet good at purely hardware-related questions and issues, making it unable to explain complex issues.
\inlinequote{Like sometimes YouTube video is even more helpful than ChatGPT when we try to like build something.}{5}. 

\vspace{-0.2cm} \paragraph{Feedback Synthesis} 
Some students utilized GenAI to reflect on their prototype's functionality and generate feedback for their peers' work, compromising the important role of reflection in the design process. Relying on GenAI for such evaluations could lead to decisions that might not align with user needs, missing the nuanced understanding that direct human engagement provides.
\inlinequote{So we can't really think of what would be a bad situation and what would be a good scenario for using our project... Yeah, well, in ideal cases, you don't want to do that.}{10}. 


%% file: 05conclusion.tex
\section{Conclusion}

\inlinequote{At the least and at most, education has changed based on AI, and allowing it in some way is necessary because straight up banning it is just impossible... Everything has to be reviewed and possibly reimagined to fit what we want students to focus their time most on}{3}

While GenAI currently enhances efficiency and fosters creativity, it also poses significant implications for students' learning experiences and outcomes in the user-centered design process. Balancing its use with deliberate, nuanced engagement in user research and feedback analysis is crucial. Adjusting HCI classes in response to GenAI's impact is essential, emphasizing the need to strengthen its empowering aspects while mitigating potential drawbacks.

%% file: 00main.bbl

\begin{thebibliography}{14}


\ifx \showCODEN    \undefined \def \showCODEN     #1{\unskip}     \fi
\ifx \showDOI      \undefined \def \showDOI       #1{#1}\fi
\ifx \showISBNx    \undefined \def \showISBNx     #1{\unskip}     \fi
\ifx \showISBNxiii \undefined \def \showISBNxiii  #1{\unskip}     \fi
\ifx \showISSN     \undefined \def \showISSN      #1{\unskip}     \fi
\ifx \showLCCN     \undefined \def \showLCCN      #1{\unskip}     \fi
\ifx \shownote     \undefined \def \shownote      #1{#1}          \fi
\ifx \showarticletitle \undefined \def \showarticletitle #1{#1}   \fi
\ifx \showURL      \undefined \def \showURL       {\relax}        \fi
\providecommand\bibfield[2]{#2}
\providecommand\bibinfo[2]{#2}
\providecommand\natexlab[1]{#1}
\providecommand\showeprint[2][]{arXiv:#2}

\bibitem[Amoozadeh et~al\mbox{.}(2024)]%
        {amoozadeh_trust_2024}
\bibfield{author}{\bibinfo{person}{Matin Amoozadeh}, \bibinfo{person}{David Daniels}, \bibinfo{person}{Daye Nam}, \bibinfo{person}{Aayush Kumar}, \bibinfo{person}{Stella Chen}, \bibinfo{person}{Michael Hilton}, \bibinfo{person}{Sruti~Srinivasa Ragavan}, {and} \bibinfo{person}{Mohammad~Amin Alipour}.} \bibinfo{year}{2024}\natexlab{}.
\newblock \bibinfo{title}{Trust in {Generative} {AI} among students: {An} {Exploratory} {Study}}.
\newblock
\newblock
\urldef\tempurl%
\url{https://doi.org/10.48550/arXiv.2310.04631}
\showDOI{\tempurl}
\newblock
\shownote{arXiv:2310.04631 [cs]}.


\bibitem[Antona et~al\mbox{.}(2023)]%
        {antona_special_2023}
\bibfield{author}{\bibinfo{person}{Margherita Antona}, \bibinfo{person}{George Margetis}, \bibinfo{person}{Stavroula Ntoa}, {and} \bibinfo{person}{Helmut Degen}.} \bibinfo{year}{2023}\natexlab{}.
\newblock \showarticletitle{Special {Issue} on {AI} in {HCI}}.
\newblock \bibinfo{journal}{\emph{International Journal of Human–Computer Interaction}} \bibinfo{volume}{39}, \bibinfo{number}{9} (\bibinfo{date}{May} \bibinfo{year}{2023}), \bibinfo{pages}{1723--1726}.
\newblock
\showISSN{1044-7318}
\urldef\tempurl%
\url{https://doi.org/10.1080/10447318.2023.2177421}
\showDOI{\tempurl}
\newblock
\shownote{Publisher: Taylor \& Francis \_eprint: https://doi.org/10.1080/10447318.2023.2177421}.


\bibitem[Baldassarre et~al\mbox{.}(2023)]%
        {baldassarre_social_2023}
\bibfield{author}{\bibinfo{person}{Maria~Teresa Baldassarre}, \bibinfo{person}{Danilo Caivano}, \bibinfo{person}{Berenice Fernandez~Nieto}, \bibinfo{person}{Domenico Gigante}, {and} \bibinfo{person}{Azzurra Ragone}.} \bibinfo{year}{2023}\natexlab{}.
\newblock \showarticletitle{The {Social} {Impact} of {Generative} {AI}: {An} {Analysis} on {ChatGPT}}. In \bibinfo{booktitle}{\emph{Proceedings of the 2023 {ACM} {Conference} on {Information} {Technology} for {Social} {Good}}}. \bibinfo{publisher}{ACM}, \bibinfo{address}{Lisbon Portugal}, \bibinfo{pages}{363--373}.
\newblock
\showISBNx{9798400701160}
\urldef\tempurl%
\url{https://doi.org/10.1145/3582515.3609555}
\showDOI{\tempurl}


\bibitem[Blythe(2023)]%
        {blythe_artificial_2023}
\bibfield{author}{\bibinfo{person}{Mark Blythe}.} \bibinfo{year}{2023}\natexlab{}.
\newblock \showarticletitle{Artificial {Design} {Fiction}: {Using} {AI} as a {Material} for {Pastiche} {Scenarios}}. In \bibinfo{booktitle}{\emph{26th {International} {Academic} {Mindtrek} {Conference}}}. \bibinfo{publisher}{ACM}, \bibinfo{address}{Tampere Finland}, \bibinfo{pages}{195--206}.
\newblock
\showISBNx{9798400708749}
\urldef\tempurl%
\url{https://doi.org/10.1145/3616961.3616987}
\showDOI{\tempurl}


\bibitem[Huang(2023)]%
        {huang_future_2023}
\bibfield{author}{\bibinfo{person}{Yuxuan Huang}.} \bibinfo{year}{2023}\natexlab{}.
\newblock \showarticletitle{The {Future} of {Generative} {AI}: {How} {GenAI} {Would} {Change} {Human}-{Computer} {Co}-creation in the {Next} 10 to 15 {Years}}. In \bibinfo{booktitle}{\emph{Companion {Proceedings} of the {Annual} {Symposium} on {Computer}-{Human} {Interaction} in {Play}}}. \bibinfo{publisher}{ACM}, \bibinfo{address}{Stratford ON Canada}, \bibinfo{pages}{322--325}.
\newblock
\showISBNx{9798400700293}
\urldef\tempurl%
\url{https://doi.org/10.1145/3573382.3616033}
\showDOI{\tempurl}


\bibitem[Muller et~al\mbox{.}(2023)]%
        {muller_genaichi_2023}
\bibfield{author}{\bibinfo{person}{Michael Muller}, \bibinfo{person}{Lydia~B Chilton}, \bibinfo{person}{Anna Kantosalo}, \bibinfo{person}{Q.~Vera Liao}, \bibinfo{person}{Mary~Lou Maher}, \bibinfo{person}{Charles~Patrick Martin}, {and} \bibinfo{person}{Greg Walsh}.} \bibinfo{year}{2023}\natexlab{}.
\newblock \showarticletitle{{GenAICHI} 2023: {Generative} {AI} and {HCI} at {CHI} 2023}. In \bibinfo{booktitle}{\emph{Extended {Abstracts} of the 2023 {CHI} {Conference} on {Human} {Factors} in {Computing} {Systems}}} \emph{(\bibinfo{series}{{CHI} {EA} '23})}. \bibinfo{publisher}{Association for Computing Machinery}, \bibinfo{address}{New York, NY, USA}, \bibinfo{pages}{1--7}.
\newblock
\showISBNx{978-1-4503-9422-2}
\urldef\tempurl%
\url{https://doi.org/10.1145/3544549.3573794}
\showDOI{\tempurl}


\bibitem[Saye(1997)]%
        {saye_technology_1997}
\bibfield{author}{\bibinfo{person}{John~W. Saye}.} \bibinfo{year}{1997}\natexlab{}.
\newblock \showarticletitle{Technology and educational empowerment: {Students}' perspectives}.
\newblock \bibinfo{journal}{\emph{Educational Technology Research and Development}} \bibinfo{volume}{45}, \bibinfo{number}{2} (\bibinfo{date}{June} \bibinfo{year}{1997}), \bibinfo{pages}{5--25}.
\newblock
\showISSN{1042-1629, 1556-6501}
\urldef\tempurl%
\url{https://doi.org/10.1007/BF02299522}
\showDOI{\tempurl}


\bibitem[Schmidt(2023)]%
        {schmidt_speeding_2023}
\bibfield{author}{\bibinfo{person}{Albrecht Schmidt}.} \bibinfo{year}{2023}\natexlab{}.
\newblock \showarticletitle{Speeding {Up} the {Engineering} of {Interactive} {Systems} with {Generative} {AI}}. In \bibinfo{booktitle}{\emph{Companion {Proceedings} of the 2023 {ACM} {SIGCHI} {Symposium} on {Engineering} {Interactive} {Computing} {Systems}}}. \bibinfo{publisher}{ACM}, \bibinfo{address}{Swansea United Kingdom}, \bibinfo{pages}{7--8}.
\newblock
\showISBNx{9798400702068}
\urldef\tempurl%
\url{https://doi.org/10.1145/3596454.3597176}
\showDOI{\tempurl}


\bibitem[Shi et~al\mbox{.}(2024)]%
        {shi_hci-centric_2024}
\bibfield{author}{\bibinfo{person}{Jingyu Shi}, \bibinfo{person}{Rahul Jain}, \bibinfo{person}{Hyungjun Doh}, \bibinfo{person}{Ryo Suzuki}, {and} \bibinfo{person}{Karthik Ramani}.} \bibinfo{year}{2024}\natexlab{}.
\newblock \bibinfo{title}{An {HCI}-{Centric} {Survey} and {Taxonomy} of {Human}-{Generative}-{AI} {Interactions}}.
\newblock
\newblock
\urldef\tempurl%
\url{http://arxiv.org/abs/2310.07127}
\showURL{%
\tempurl}
\newblock
\shownote{arXiv:2310.07127 [cs]}.


\bibitem[Shin et~al\mbox{.}(2023)]%
        {shin_integrating_2023}
\bibfield{author}{\bibinfo{person}{Joon~Gi Shin}, \bibinfo{person}{Janin Koch}, \bibinfo{person}{Andrés Lucero}, \bibinfo{person}{Peter Dalsgaard}, {and} \bibinfo{person}{Wendy~E. Mackay}.} \bibinfo{year}{2023}\natexlab{}.
\newblock \showarticletitle{Integrating {AI} in {Human}-{Human} {Collaborative} {Ideation}}. In \bibinfo{booktitle}{\emph{Extended {Abstracts} of the 2023 {CHI} {Conference} on {Human} {Factors} in {Computing} {Systems}}} \emph{(\bibinfo{series}{{CHI} {EA} '23})}. \bibinfo{publisher}{Association for Computing Machinery}, \bibinfo{address}{New York, NY, USA}, \bibinfo{pages}{1--5}.
\newblock
\showISBNx{978-1-4503-9422-2}
\urldef\tempurl%
\url{https://doi.org/10.1145/3544549.3573802}
\showDOI{\tempurl}


\bibitem[Wang et~al\mbox{.}(2019)]%
        {wang_human-ai_2019}
\bibfield{author}{\bibinfo{person}{Dakuo Wang}, \bibinfo{person}{Justin~D. Weisz}, \bibinfo{person}{Michael Muller}, \bibinfo{person}{Parikshit Ram}, \bibinfo{person}{Werner Geyer}, \bibinfo{person}{Casey Dugan}, \bibinfo{person}{Yla Tausczik}, \bibinfo{person}{Horst Samulowitz}, {and} \bibinfo{person}{Alexander Gray}.} \bibinfo{year}{2019}\natexlab{}.
\newblock \showarticletitle{Human-{AI} {Collaboration} in {Data} {Science}: {Exploring} {Data} {Scientists}' {Perceptions} of {Automated} {AI}}.
\newblock \bibinfo{journal}{\emph{Proceedings of the ACM on Human-Computer Interaction}} \bibinfo{volume}{3}, \bibinfo{number}{CSCW} (\bibinfo{date}{Nov.} \bibinfo{year}{2019}), \bibinfo{pages}{211:1--211:24}.
\newblock
\urldef\tempurl%
\url{https://doi.org/10.1145/3359313}
\showDOI{\tempurl}


\bibitem[Xiao et~al\mbox{.}(2023)]%
        {xiao_supporting_2023}
\bibfield{author}{\bibinfo{person}{Ziang Xiao}, \bibinfo{person}{Xingdi Yuan}, \bibinfo{person}{Q.~Vera Liao}, \bibinfo{person}{Rania Abdelghani}, {and} \bibinfo{person}{Pierre-Yves Oudeyer}.} \bibinfo{year}{2023}\natexlab{}.
\newblock \showarticletitle{Supporting {Qualitative} {Analysis} with {Large} {Language} {Models}: {Combining} {Codebook} with {GPT}-3 for {Deductive} {Coding}}. In \bibinfo{booktitle}{\emph{Companion {Proceedings} of the 28th {International} {Conference} on {Intelligent} {User} {Interfaces}}} \emph{(\bibinfo{series}{{IUI} '23 {Companion}})}. \bibinfo{publisher}{Association for Computing Machinery}, \bibinfo{address}{New York, NY, USA}, \bibinfo{pages}{75--78}.
\newblock
\showISBNx{9798400701078}
\urldef\tempurl%
\url{https://doi.org/10.1145/3581754.3584136}
\showDOI{\tempurl}


\bibitem[Yang et~al\mbox{.}(2020)]%
        {yang_re-examining_2020}
\bibfield{author}{\bibinfo{person}{Qian Yang}, \bibinfo{person}{Aaron Steinfeld}, \bibinfo{person}{Carolyn Rosé}, {and} \bibinfo{person}{John Zimmerman}.} \bibinfo{year}{2020}\natexlab{}.
\newblock \showarticletitle{Re-examining {Whether}, {Why}, and {How} {Human}-{AI} {Interaction} {Is} {Uniquely} {Difficult} to {Design}}. In \bibinfo{booktitle}{\emph{Proceedings of the 2020 {CHI} {Conference} on {Human} {Factors} in {Computing} {Systems}}} \emph{(\bibinfo{series}{{CHI} '20})}. \bibinfo{publisher}{Association for Computing Machinery}, \bibinfo{address}{New York, NY, USA}, \bibinfo{pages}{1--13}.
\newblock
\showISBNx{978-1-4503-6708-0}
\urldef\tempurl%
\url{https://doi.org/10.1145/3313831.3376301}
\showDOI{\tempurl}


\bibitem[Zheng(2023)]%
        {zheng_chatgpt_2023}
\bibfield{author}{\bibinfo{person}{Yong Zheng}.} \bibinfo{year}{2023}\natexlab{}.
\newblock \showarticletitle{{ChatGPT} for {Teaching} and {Learning}: {An} {Experience} from {Data} {Science} {Education}}. In \bibinfo{booktitle}{\emph{The 24th {Annual} {Conference} on {Information} {Technology} {Education}}}. \bibinfo{publisher}{ACM}, \bibinfo{address}{Marietta GA USA}, \bibinfo{pages}{66--72}.
\newblock
\showISBNx{9798400701306}
\urldef\tempurl%
\url{https://doi.org/10.1145/3585059.3611431}
\showDOI{\tempurl}


\end{thebibliography}
